\documentclass[doublecol]{epl2} 

\usepackage{mymcite}

\usepackage{amsmath}
\usepackage{amsfonts}
\usepackage{amssymb}

\usepackage{graphicx}
\usepackage{dcolumn}
\usepackage{bm}
\usepackage[final,dvips]{epsfig}
\usepackage{bm}
\usepackage{bbm}
\usepackage{color}
\usepackage{verbatim}
\usepackage{hyperref}

\newcommand{\be}{\begin{equation}}
\newcommand{\ee}{\end{equation}}
\newcommand{\ba}{\begin{eqnarray}}
\newcommand{\ea}{\end{eqnarray}}

\newcommand{\vv}[1]{{\mathbf #1}}
\newcommand{\vvv}[1]{\mbox{\boldmath{$#1$}}}
\newcommand{\qaf}{{\vv q_{AF}}}

\newcommand{\each}[1]{ #1 }

\title{ 
Theory of two-particle excitations and the magnetic
  susceptibility in high-T$_c$ cuprate superconductors
} 
\shorttitle{Two-particle excitations in high-T$_c$ superconductors}

\author{S. Brehm\inst{1}, E. Arrigoni\inst{2}, M. Aichhorn\inst{3},
  and W. Hanke\inst{1} }
\shortauthor{S. Brehm \etal}

\institute{                    
\inst{1} Institute for Theoretical Physics and Astrophysics, University of
W\"urzburg, Am Hubland, 97074 W\"urzburg, Germany\\
\inst{2} Institute of Theoretical Physics and Computational
Physics, Graz University of Technology, Petersgasse 16, 8010 Graz, Austria\\
\inst{3} Centre de Physique Th\'{e}orique, \'{E}cole Polytechnique,
91128 Palaiseau Cedex, France 
}

\date{29/10/2009}


\abstract{
  Two-particle (2-p) excitations such as spin and charge
  excitations play a key role in high-T$_c$ cuprate superconductors
  (HTSC). On the basis of a parameter-free theory, which extends
  the Variational Cluster Approach (a recently developed
  embedded cluster method) to 2-p excitations, the magnetic
  excitations of HTSC are shown to be reproduced for
  a Hubbard model  within the relevant strong-coupling regime. In
  particular, the resonance mode in the underdoped regime, its
  intensity and ``hour-glass'' dispersion are in good overall agreement
  with experiments.
}

\begin{document}

\maketitle

\section{Introduction}

Two-particle (2-p) excitations and their corresponding magnetic,
charge, optical and pairing susceptibilities are fundamental for
obtaining a microscopic understanding of the 
high-T$_c$ cuprate superconductor (HTSC) physics, complementing
single-particle (such as angle-resolved photoemission (ARPES),
etc.) experiments. A key example is provided by the magnetic excitations:
When entering the superconducting (SC) state in the
$\textrm{high-T}_\textrm{c}$ cuprates, the magnetic excitation
spectrum is characteristically and markedly modified: a resonant mode
emerges with its peak intensity being highest around the wave vector
$\qaf=(\pi,\pi)$ characteristic of antiferromagnetism (AF) in the undoped parent compound \cite{ro.re.91,*ar.ni.99,*re.bo.04,*si.pa.07,*hi.bo.07,pa.si.04}.
Its frequency $\omega_{\textrm{res}}(\qaf)$ follows the doping
dependence of $\textrm{T}_\textrm{c}$. Away from
$\qaf$, the mode has both a downward and upward
``hour-glass''-like dispersion. A variety of experiments in the
HTSC, such as photoemission, optical and tunneling spectroscopies, have been
interpreted as evidence of interactions of electrons with this
mode \cite{see2,*bo.ko.03,*hw.ti.04,*za.oz.01}. However its microscopic origin, in particular its role in
pairing and the more detailed effects arising from the interactions of
charge carriers with this magnetic mode are still unclear and
intensively debated \cite{see3,*vo.vo.06,*uh.sc.04,*ke.ki.02,*ab.ch.02,*norm.01,*esch.06}. A prerequisite to resolve this debate obviously
requires a consistent theoretical description of the neutron
resonance mode and, more generally, the magnetic excitation spectrum \cite{ka.bi.98,*ke.bi.91,*ke.be.92}
{\it and} at the same time of the phase diagram, containing the competing AF and
SC phases.

In this letter, on the basis of a microscopic theory for 2-p excitations,
we provide such a consistent description for the experimentally
relevant regime of the two-dimensional (2D) Hubbard
model. 
The essential new points here are that 
our theory for
2-p excitations (e.g. the dynamic spin-susceptibility) is (i)
parameter-free (given fixed, widely-accepted values for the Hubbard model parameters)
and (ii) is working in the relevant strong correlation regime
of the underlying Hubbard
model. Previous descriptions of the magnetic resonance have been
obtained by weak-coupling \cite{fora,*er.mo.05} and/or semiphenomenological approaches \cite{see,*ab.ch.99,*se.pr.03,*pr.se.06,*zeyh.08u,de.zh.95,*de.ha.04}
reproducing the experimental behavior with adjustable parameters. 
An accurate description of the infinite lattice 
is crucial \each{ in order } to obtain the magnetic resonance which may be
considered as a ``fingerprint'' of the AF order in the SC state. Only
then are we able to differentiate between the competing AF and SC orders in the
phase diagram. Therefore, 
\each{ the infinite-lattice limit (not to be confused with the limit
  of infinite dimensions) }
 has also to be embedded in a
controlled description of the corresponding susceptibilities.

We extend the original idea of the Variational Cluster Approach (VCA), 
which is to extrapolate cluster results to the 
\each{infinite lattice,}
to the treatment of 2-p excitations.
In our novel approach, the 2-p vertex extracted from the
corresponding cluster susceptibilities is used to obtain the
susceptibilities 
\each{for an infinite lattice. }
The VCA was recently
applied to calculate the zero-temperature (T=0) phase diagram as well
as single-particle excitations \cite{po.ai.03,*da.ai.04,se.la.05,ai.ar.05,*ai.ar.06,*ai.ar.07.ps,*ai.ar.06.vc}
of the single-band Hubbard model. These results succesfully reproduced salient
experimental features such as the electron-hole asymmetry in the
doping dependence of AF and SC phases \cite{se.la.05,ai.ar.05,*ai.ar.06,*ai.ar.07.ps,*ai.ar.06.vc} in the HTSC materials.
Also the VCA single-particle excitations were found to reproduce characteristic
features observed in ARPES experiments.
\each{
In particular, in Ref. \cite{ai.ar.07},
the  magnitude and doping dependence of the SC gap near the nodal and
antinodal regions was studied in detail, and shown to reproduce qualitatively
the much-discussed
presence of a gap dichotomy of the nodal and antinodal SC gaps.
}
 Combined with the new results for 2-p magnetic
excitations, presented in this work, a consistent picture emerges,
which lends substantial support to Hubbard-model descriptions of
$\textrm{high-T}_\textrm{c}$ cuprate superconductivity.

For the appropriate strongly correlated regime ($U=8t$) of the
underdoped Hubbard model the resonance is obtained in
a parameter-free calculation and verified to be a spin $S=1$ excitonic bound state, which appears in the
SC-induced gap in the spectrum of electron-hole spin-flip (i.e. $S=1$)
excitations. This will be detailed in our results, where we find the doping dependence of
$\omega_{\textrm{res}}(\qaf)$, the energy-integrated
spectral weight evaluated at $\qaf$ and the difference of the magnetic susceptibilities in the SC and the normal
(N) states to be in qualitative accord with neutron scattering data for underdoped
$\textrm{YBa}_{\textrm{2}}\textrm{Cu}_{\textrm{3}}\textrm{O}_{\textrm{6+x}}$
(YBCO), where the mode was studied in great detail \cite{ro.re.91,*ar.ni.99,*re.bo.04,*si.pa.07,*hi.bo.07,pa.si.04}.
A spin excitonic bound state has previously been suggested on the basis of an
itinerant picture, most frequently invoking a weakly correlated
RPA-like form of the dynamic spin susceptibility (for a recent
reference see, for example, Ref.~\cite{fora,*er.mo.05}). As a
weak-coupling form it leads to a Fermi-liquid like $\chi(q,\omega)$,
which is in contrast to some of the anomalous dynamics found in
neutron scattering experiments \cite{ro.re.91,*ar.ni.99,*re.bo.04,*si.pa.07,*hi.bo.07,pa.si.04}. On the other hand, when
the 2-p interaction and the SC gap are used as adjustable
parameters, it qualitatively accounts for the mode behavior near optimal and
overdoped regimes \cite{fora,*er.mo.05}.

\section{Model and Method}
We start from the two-dimensional Hubbard model:
\be
\label{eq:hubbard}
H = - \sum_{ij \sigma} t_{ij} c_{i\sigma}^\dagger c_{j\sigma} 
+ U \sum_i n_{i\uparrow} n_{i\downarrow} \ ,
\ee
where $t_{ij}$ denote nearest ($t$) neighbor and 
next-nearest ($t'=-0.3 t$) neighbor hopping matrix elements, $c_{i\sigma}^\dagger$ and $c_{j\sigma}$ are the usual creation
and destruction operators, $n_{i\sigma}$ their density and $U=8t$ the local Hubbard repulsion.

We consider the following matrix expression for the transverse spin susceptibility:
\be
\label{eq:bethe-sal}
\vvv \chi(\vv Q, i \omega_m^b)=\vvv \chi^0(\vv Q, i
\omega_m^b)+
\ee
\be
\nonumber
+\vvv \chi^0(\vv Q, i \omega_m^b) \vvv \Gamma(\vv Q, i
\omega_m^b) \vvv \chi(\vv Q, i \omega_m^b) \ ,
\ee
with the bosonic Matsubara frequencies $\omega^b_m=2 m \pi
T$ and T the temperature. 

Within our embedded cluster approach, the susceptibility $\bm{\chi}$
depends on the wave vector $\vv Q$ in the reduced Brillouin zone (BZ)
associated with the superlattice produced by the clusters. In
addition, $\bm{\chi}$ is considered as a matrix in the
cluster-site indices $i$ and $j$ and the products in
Eq.~(\ref{eq:bethe-sal}) are matrix products. 
Its definition is given by
\be
\label{eq:chi0}
\chi_{ij}(\vv Q, i \omega_m^b)\equiv
\int^{\beta}_0 d \tau \ e^{i \omega^b_m
  \tau} \langle
S^-_{\vec{i}}(\vv Q,\tau) S^+_{\vec{j}}(-\vv Q,0) \rangle,
\ee
with the mixed representation for the spin operator
\be
\label{mix}
S^{a}_{\vec{i}}(\vv Q,\tau) = 
\frac{1}{N_{cl}} \sum_{\vv R} S^{a}_{\vec{i}+\vv R}(\tau)
e^{i\vv Q \cdot \vv R} \;.
\ee
Here, the sum is carried out over all clusters, whose
position is
given by the vector $\vv R$, 
$N_{cl}$ is the number of clusters, $S^{a}_{\vec i+\vv R}$ is the $a$
component of the spin operator at site ${\vec i+\vv R}$ and $\tau$ is 
the usual imaginary time.

In Eq.~(\ref{eq:bethe-sal}), an effective particle-hole interaction (matrix)
$\bm{\Gamma}$ is introduced, obtained from an average of the 2-p vertex
over the additional internal frequencies and momenta~\footnote{
  Notice that, besides this approximation, Eq.~(\ref{eq:bethe-sal}) is
  the well-known Bethe-Salpeter equation, which is in principle exact and valid
  for arbitrary interaction and is not limited to weak coupling as the RPA
  approximation. Neglecting the explicit dependence of the internal
  momenta and frequencies has been found to be a reasonable
  approximation in finite-T single-cluster QMC calculations
  \cite{bu.sc.95} and in an extension of the Dynamical Cluster Approximation (DCA) to 2-p
  susceptibility calculations \cite{ho.as.08}. However, we would
  like to stress that, while the Eq.~(\ref{eq:bethe-sal}) appears
  formally similar to the equation used in Ref.~\cite{ho.as.08} for
  DCA 2-p quantities, subtle 
  differences occur in building up
  the self-consistent (spin-) response within a cluster: This is due to
  the fact that in the DCA (in contrast to our present VCA-type scheme)
  each cluster site ``sees'' the same mean-field. We have found, as shown
  e.g. in the ``consistency number'' $\alpha$
  (Eq.~(\ref{eq:gamma-cluster},\ref{eq:alpha-sum})), the present
  scheme to give significantly improved results for comparable cluster sizes.}
Eqs.(\ref{eq:bethe-sal}-\ref{mix}) as well as the following
discourse can straightforwardly be generalized to other 2-p
susceptibilities, such as the charge response function, by replacing the spin operator in
Eq.~(\ref{eq:chi0}) by the corresponding charge density operator.

The ``bubble'' susceptibility  $\vvv \chi^0$ in \eqref{eq:bethe-sal}
is obtained as a convolution of the ``dressed'' VCA one-particle (1-p) Green's functions, i.e.
\be
\nonumber
\chi^0_{ij}(\vv Q, i \omega_m^b)= - \frac{T}{N_{cl}} \sum_{n, \vv K}
\ee
\be
\label{chi0}
\Big(
G^{VCA}_{ij\uparrow}(\vv K+\vv Q,i\omega_n^f+i\omega_m^b) G^{VCA}_{ji\downarrow}(\vv K,i\omega_n^f)+
\ee
\be
\nonumber   
+F^{VCA}_{ij}(\vv Q-\vv K,i\omega_m^b-i\omega_n^f) {F^*}^{VCA}_{ji}(\vv K,i\omega_n^f)
\Big) \ {.}
\ee
Here, $i \omega_n^f=(2n+1)\pi T$ denote fermionic Matsubara frequencies, 
$G^{VCA}$ are normal and  $F^{VCA}$ anomalous (i. e. characteristic of the SC state) fully interacting
Green's functions obtained within the VCA and expressed in a mixed representation
analogous to \eqref{mix}, and $\vv K$ is again a vector of the reduced BZ.

Within the VCA \cite{po.ai.03,*da.ai.04}, the self-energy of a so-called ``reference system'' is used as an
approximation to the one of the physical system. The
reference system  typically consists of a subcluster of the
original system in which single-particle terms can be ``optimized'' to obtain 
a stationary point of the grand potential for the 
infinite lattice.
In  a similar spirit, the effective interaction $\vvv \Gamma$ in 
Eq.~(\ref{eq:bethe-sal}) is obtained from the corresponding cluster quantity, 
\be
\label{eq:gamma-cluster}
\vvv\Gamma(i \omega_m^b) = \alpha \left[(\vvv
  \chi_{\textrm{cluster}}^0(i \omega_m^b))^{-1} - (\vvv
  \chi_{\textrm{cluster}}(i \omega_m^b))^{-1} \right]
\ee
(cf. also Ref.~\cite{ja.ma.01}). Note, that this vertex is defined on
the cluster and thus independent of $\vv Q$.
In Eq.~(\ref{eq:gamma-cluster}), $\vvv \chi_{\textrm{cluster}}^0$ is
calculated via the same convolution as in \eqref{chi0}, but using
the exact single-particle cluster Green's functions, and $\vvv \chi_{\textrm{cluster}}$ is
the exact cluster susceptibility. The latter quantities are obtained
via Lanczos exact diagonalization.

In Eq.~(\ref{eq:gamma-cluster}), we have introduced a multiplicative
constant $\alpha$, which is \textit{not} a free parameter, but rather
it is fixed by enforcing the sum rule for the transverse spin susceptibility:
\be
\frac{T}{N_{cl}}\sum_{\vv Q,i \omega_m^b}\chi_{ii}(\vv Q,i \omega_m^b)=\langle S^-_i S^+_i \rangle
\label{eq:alpha-sum}
\ee
In our parameter-free approach, the value of $\alpha$ can
be used to assess the quality of our 2-p scheme:
when $\alpha$ is close to $1$, the cluster vertex is a good
approximation to the vertex 
of the infinite \each{ lattice }
 (see Fig.~\ref{fig1}b).
The quality of this approximation obviously
depends on the cluster size $L_c$, and is expected to
become exact for $L_c \rightarrow \infty$, where also $\alpha
\rightarrow 1$. It also  certainly depends on the ratio $\frac{t}{U}$ and on
doping. We expect the method to give better results at
strong coupling, where short-range effects dominate and are, thus, well
accounted for by modest cluster sizes. Below, we provide some results in support
of this expectation. After Fourier transforming over the cluster
sites, the spin susceptibility $\chi(\vv Q + \vv k, \vv Q + \vv k', i \omega_m^b)$ 
acquires a dependence on two momenta due to the translation symmetry breaking introduced by
the cluster tiling. The {\em physical} translation-invariant susceptibility $\chi(\vv q,i \omega_m^b)$, with $\vv q = \vv
Q+\vv k$ is taken to be its diagonal part $\chi(\vv Q + \vv k, \vv Q + \vv k, i \omega_m^b)$.

\section{Results}
\begin{figure}[t]
  \includegraphics[width=8cm]{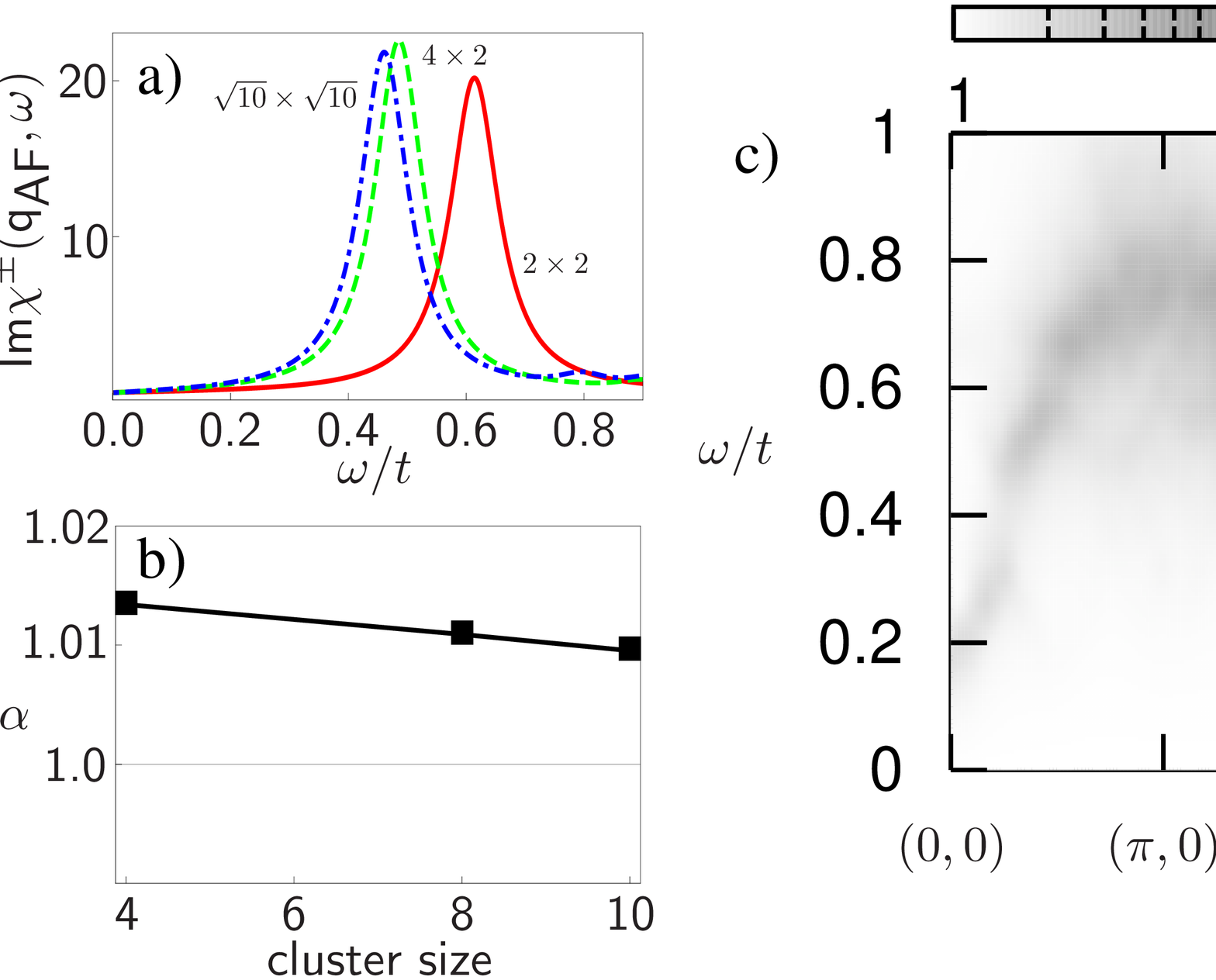}
  \caption{
    VCA calculation for the 2D Hubbard model $(U=8t)$: (a) The imaginary part of the susceptibility
    $\chi^{\pm}(\qaf,\omega)$ \each{at half filling} and (b) the sum-rule constant $\alpha$ as a function of the cluster size at
    half-filling; (c) $\textrm{Im}\chi^{\pm}(\bm{q},\omega)$ intensity
    plot at \each{$4\%$} hole doping.
  }
\label{fig1}
\end{figure}
Fig.~\ref{fig1}a displays our results for the $\vv q=\qaf$ part of the transverse
spin spectral function $\textrm{Im}\chi^{\pm}(\bm{q}=\qaf,\omega)$ at
half-filling, i.e. in the AF phase, for different cluster sizes
$L_c=2\times2$, $4\times2$ and $\sqrt{10}\times\sqrt{10}$. Here,
a ``finite-size'' gap appears ,
\each{ which 
 continuously } diminishes
with increasing cluster sizes (Fig.~\ref{fig1}a). It is comforting to note that the
controlling constant $\alpha$ is very close to and also continuously decreasing towards
$\alpha=1$ as a function of increasing cluster size (Fig.~\ref{fig1}b).

Fig.~\ref{fig1}c shows the corresponding intensity plot for $\textrm{Im}
\chi^{\pm}(\bm{q},\omega)$ with remnants of the spin-density wave
dispersion for a 
\each{ $L_c=2\times4$}
 ``reference''
cluster at \each{4\%} hole doping. Results obtained with different cluster
sizes reveal that at finite doping ``finite-size'' effects are of minor
importance compared to the half-filled situation. We attribute this to
a screening effect, which renders the 2-p vertex $\Gamma$
significantly more short-ranged, i.e. more local. This means that it
can accurately be extracted from the exact diagonalization of relatively
small clusters.
\begin{figure}[t]
  \includegraphics[width=8cm]{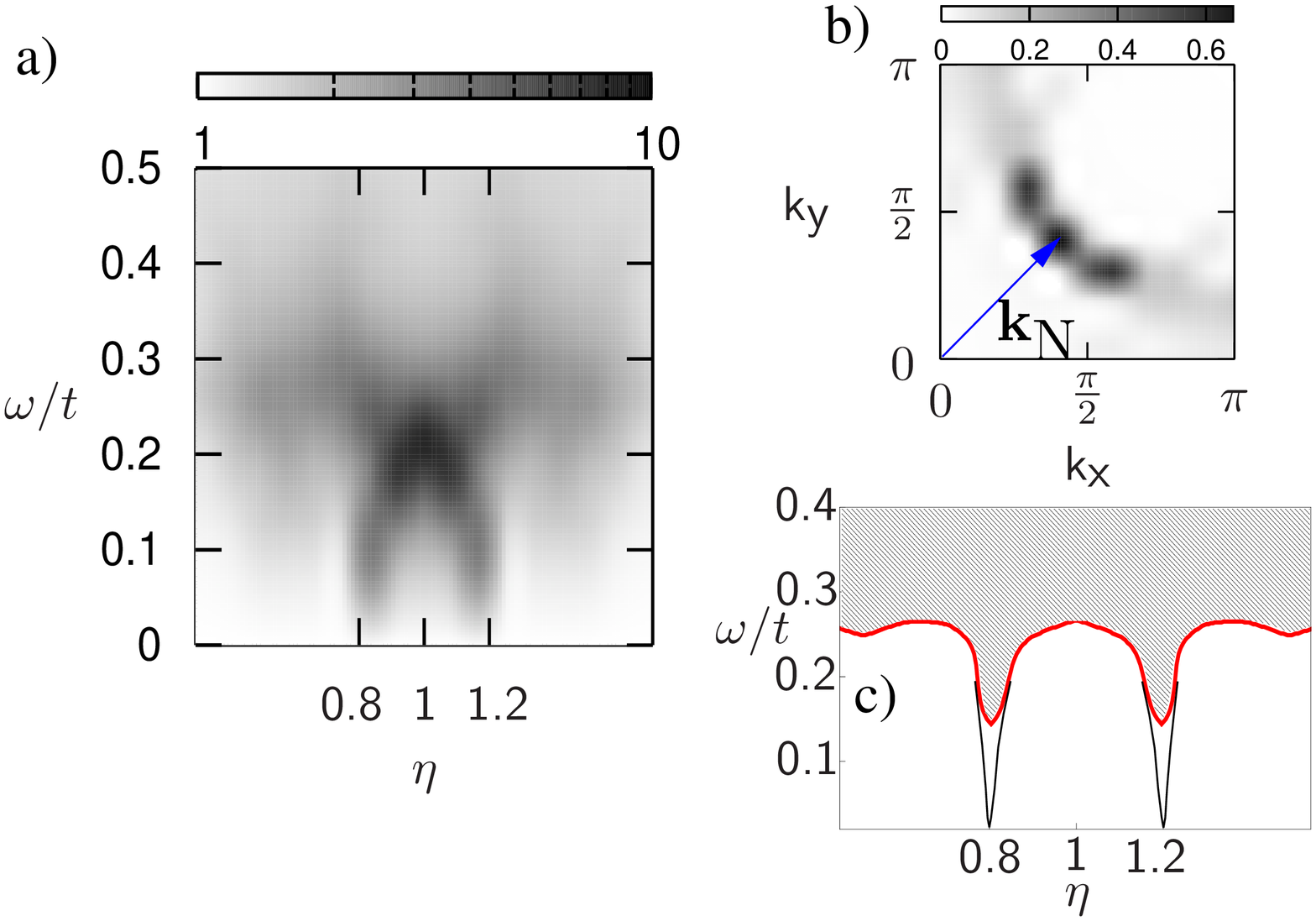}
  \caption{ (a) Intensity plot of
    $\textrm{Im}\chi^{\pm}(\bm{q},\omega)$ from the microscopic 2-p
    theory Eqs.~(\ref{eq:bethe-sal}-\ref{eq:alpha-sum}) along $\bm{q}=\eta (\pi,\pi)$ 
    displaying the ``hour-glass'' shape. (b) Intensity plot of the
    low-energy spectral weight obtained from the corresponding VCA
    calculation for the 1-p spectral weight displaying the Fermi
    surface with the nodal scattering vector $2
    \bm{k}_{\textrm{N}}\simeq 0.8 (\pi,\pi)$. (c) Spin-flip
    electron-hole continuum (hatched area: extracted from
    Eq.~(\ref{eq:chi0})) with a minimum at $2
    \bm{k}_{\textrm{N}}$. All results are obtained in the SC phase at $x=0.18$.}
\label{fig2}
\end{figure}

Our results for the magnetic response properties in the SC state are shown
in Fig.~\ref{fig2}. Fig.~\ref{fig2}a displays a density plot of the spin spectral
function at x=0.18 doping ($3 \times 3$ ``reference'' cluster). The
magnetic resonance emerges in the SC-induced gap of $S=1$ electron-hole (e-h) excitations
when entering this SC doping regime. In Fig.~\ref{fig2}a, we plot the intensity in
the $(\omega, \bm{q})$-plane along the diagonal of the 2D-BZ. The
celebrated ``hour-glass'' structure observed in the experiments (see,
in particular, Ref.~\cite{pa.si.04}) is very well reproduced in our calculation.
The structure has its maximum spectral weight confined to a region
close to $\qaf$ and a dramatic intensity reduction around $\simeq 0.8 (\pi,\pi)$.

Our results confirm the experimental interpretation put forward in
Ref.~\cite{pa.si.04}. As shown {in a density plot} in Fig.~\ref{fig2}b,  we
find for the  same  doping as  in Fig.~\ref{fig2}a ($x=0.18$) a typical Fermi
surface closed around $(\pi,\pi)$. Fig.~\ref{fig2}c plots the corresponding e-h
continuum (i.e. $S=1$, spin-flip e-h excitations), obtained
in our VCA calculation. Only collective modes below the e-h
continuum (red line) can actually be detected, because modes within the continuum
are Landau damped. This e-h continuum corresponds to Fig.~{4c} in
Ref.~\cite{pa.si.04}. The continuum threshold exhibits also in our
case a pronounced minimum in the vicinity of the wave vector $2 \bm{k}_\textrm{N}\simeq 0.8 (\pi,\pi)$,
which corresponds to scattering between nodes of the d-wave gap function (Fig.~\ref{fig2}b gives just one quadrant of the BZ). The
minimum in our calculation is, however, not so steep as in the idealistic situation in Fig.~4c of Ref.~\cite{pa.si.04},
due to correlation effects  and to a broadening of {$0.05t$} used for the exact diagonalization.

We would like to emphasize that a similar picture has been suggested
in RPA-like descriptions of the neutron resonance (see, for example, Ref.~\cite{fora,*er.mo.05}). 
However, in these calculations the d-wave gap amplitude as well as the
magnitude of the effective 2-p interaction have been
introduced as adjustable parameters in order to reproduce the experimental energy positions of
the resonance mode at $(\pi,\pi)$ and the e-h threshold around $0.8(\pi,\pi)$. 
\begin{figure}[t]
  \includegraphics[width=8cm,height=9cm]{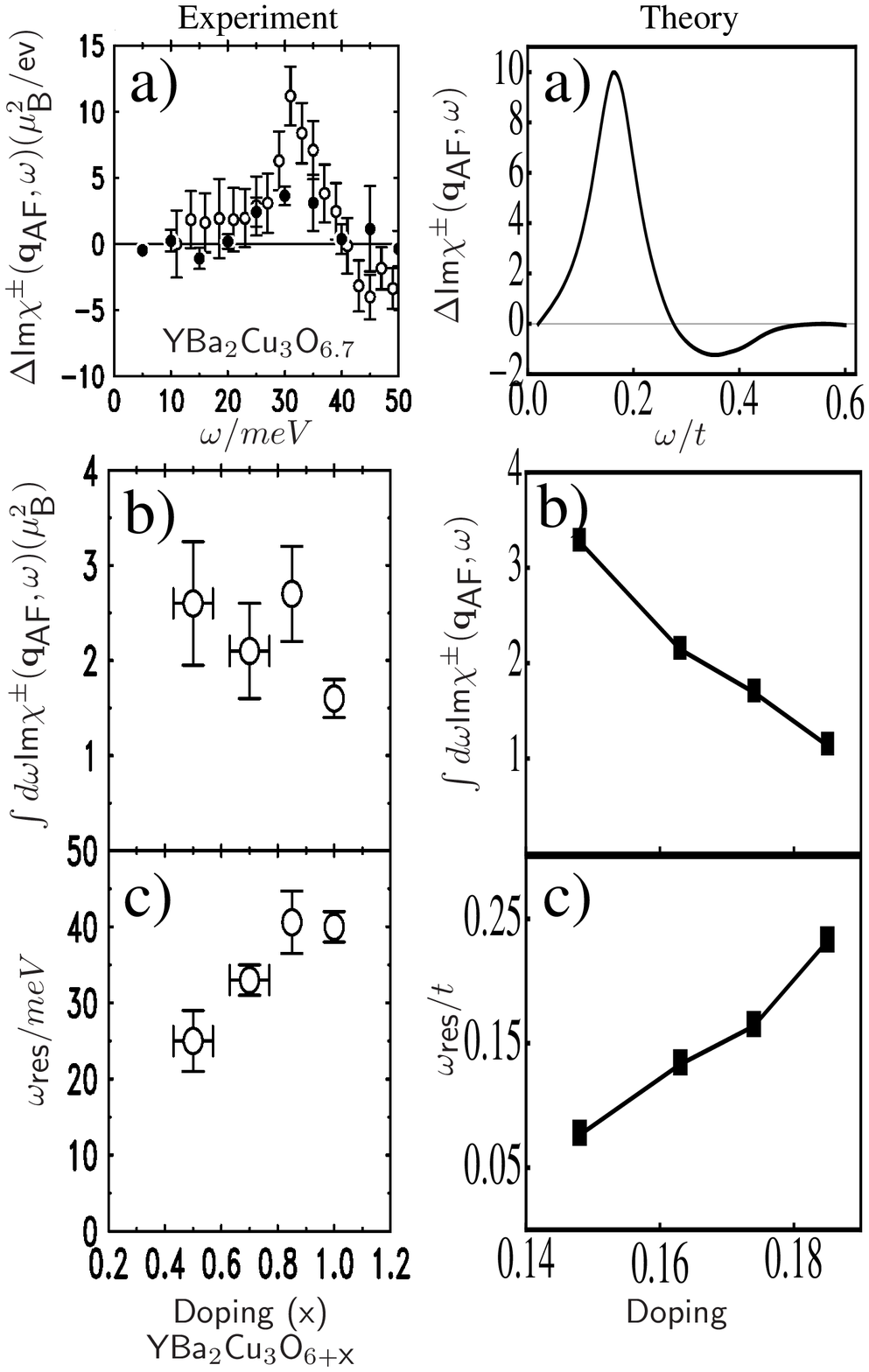}
  \caption{
    Comparisons of our theoretical results with 
    experiment (reprinted with kind permission from
    Ref.~\href{http://prola.aps.org/abstract/PRB/v61/i21/p14773_1}{\protect\cite{fo.bo.00}}
    ``Copyright (2000) by the American Physical Society''):
    (a) Difference between
    $\textrm{Im}\chi(\qaf,\omega)$ in the SC and
    normal states 
    (theory: $x=0.17$); (b) $\omega$-integrated spectral weight at
    $\qaf$; (c) $\omega_{\textrm{res}}$ as a
    function of doping.}
\label{fig3}
\end{figure}

In Figs.~\ref{fig3}a to \ref{fig3}c, we show additional comparisons of our calculations with salient features of the neutron
scattering experiments in underdoped
$\textrm{YBa}_{\textrm{2}}\textrm{Cu}_{\textrm{3}}\textrm{O}_{\textrm{6+x}}$
\cite{fo.bo.00}: Fig.~\ref{fig3}a displays the difference ($\textrm{Im} \Delta
\chi^{\pm}(\qaf,\omega)$) between the spin spectral functions in the SC and in the N phases.
The neutron results  by Fong et al. \cite{fo.bo.00} (left panel) are compared with our results (right
panel). Since we are using a $T=0$ method, our ``normal-state''
solutions have been obtained by forcing the SC Weiss field to be zero in the variational
procedure. Furthermore, the dopings in the experimental and in the
theoretical curves do not exactly coincide, although both are in the underdoped region.
In this sense, our comparison is only qualitative. Nevertheless,  our calculations reproduce the experimental
finding, that the enhancement of the spectral weight around the
resonance peak energy is accompanied by a reduction of the spectral
weight over a limited energy range both above and below $\omega_{\textrm{res}}(\qaf)$. 

Fig.~\ref{fig3}b compares the energy-integrated spin spectral weight at $\qaf$
obtained in experiment at various dopings \cite{fo.bo.00} (left panel)
with our theoretical results (right) in the SC region. The overall doping dependence is similar. 
Finally, in Fig.~\ref{fig3}c, the doping behavior of the energy of the magnetic
resonance is compared in the underdoped regime ($x=1$ corresponds in
experiment to optimal doping). Again a similar trend is observed.

\section{Summary and conclusions}
In summary, our novel theory for 2-p excitations is able
to provide an appropriate description of the resonance mode in HTSC in
good agreement with experiments. In particular, the calculated doping
dependence of $\omega_{\textrm{res}}(\qaf)$, the ``hour-glass''
dispersion of the resonance and its rapid decrease around
a characteristic wave vector $2 \bm{k}_\textrm{N}$, which coincides
with the distance between nodal points on the Fermi surface, are
qualitatively consistent with the
experiment and support the $S=1$ magnetic exciton scenario.
In contrast to previous calculations, our results are obtained 
in the appropriate strong-correlation regime and contain no
adjustable parameters.  When taken together  with earlier
results on the phase diagram and single-particle excitations,
they constitute a rather strong support for {a} Hubbard-model description of
the HTSC materials.

\acknowledgments

It is a pleasure to thank M.~Potthoff, D.J.~Scalapino and S.~Hochkeppel for discussions.
The work is supported by the Deutsche Forschungsgemeinschaft within the 
Forschergruppe FOR~538 and by the Austrian Science Fund (FWF), grants P18551-N16 and J2760-N16.


\end{document}